\documentclass[runningheads]{svmult}

\usepackage{makeidx}   
\usepackage{graphicx}  
\usepackage{subeqnar}  
\usepackage{multicol}  
\usepackage{physprbb}  
\makeindex             

\begin{document}
\title*{Mass and Angular Momentum of Sgr\,A*}
\toctitle{Mass and Angular Momentum of Sgr\,A*}
\titlerunning{Mass and Angular Momentum of Sgr\,A*}
%
\author{Bernd Aschenbach}
\authorrunning{Bernd Aschenbach}
%
%
\institute{Max-Planck-Institut f\"ur extraterrestrische Physik, 
           Garching, 85740, Germany  }

\maketitle              

\noindent Over the past four years it has become increasingly clear that Sgr\,A* is 
not a steady source but is more or less continuously flaring. 
The first large flare was observed in X-rays on October 26, 2000 with 
{\it{Chandra}} [1]. 
A second, even brighter X-ray flare went off on October 3, 2002 and was 
recorded by {\it{XMM-Newton}} [2]. 
In the near infrared (NIR) two bright flares were observed on June 15 and 16, 
2003 with the VLT {\it{NACO}} [3]. The two NIR flares showed 
quasi-periodic oscillations with a period of 16.8$\pm$2 min which if 
interpreted as the Kepler period of the last, marginally stable orbit implies 
a spin of the Sgr\,A* black hole (BH) of $a = 0.52$ adopting  a BH mass of 
3.6$\times$10$\sp 6 \rm{M\sb{\odot}}$ [3]. The two X-ray flares did not only show 
a quasi-period consistent with the NIR period but indicated additional 
quasi-periods, which fall into four groups [4] (c.f. Fig. 1). 
In a first attempt we associated 
these quasi-periods with the three fundamental oscillation modes a particle 
can have when orbiting a rotating BH representative for an 
accretion disk [4]. These are the Kepler frequency $\rm{\Omega\sb K}$ (azimuthal), the radial epicyclic 
frequency $\rm{\Omega\sb R}$ (radial) and the vertical epicyclic frequency $\rm{\Omega\sb V}$ (polar). 
For the first three quasi-periods a consistent solution 
was found for $\rm{M\sb{BH}} = 2.72\sp{+0.12}\sb{-0.19}\times$10$\sp 6$ $\rm{M\sb{\odot}}$ 
and $a =0.9939\sp{+0.0026}\sb{-0.0074}$. However, it had to be assumed that the oscillations 
originate from two different orbit radii, and for the fourth period no satisfactory explanation 
was found [4].

\begin{figure}[t]
\begin{center}
\includegraphics[bb=28pt 65pt 566pt 608pt,angle=-90,width=.78\textwidth]{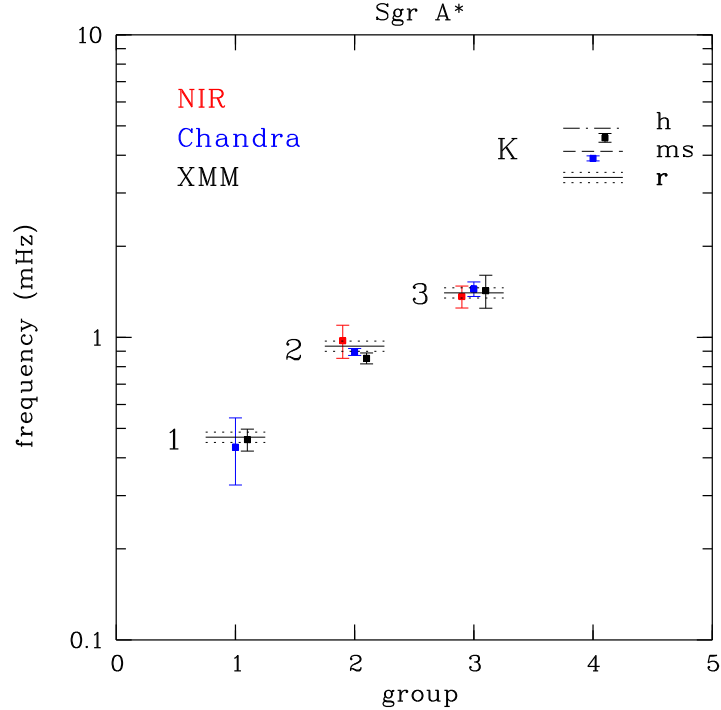}
\end{center}
\caption[]{The four groups of frequencies found in the NIR and X-ray flares of Sgr\,A*. The solid horizontal
lines indicate the best fit to the measured frequencies of groups 1, 2 and 3 for a frequency ratio of 1:2:3; the
associated dashed lines correspond to $\pm$ 1$\rm{\sigma}$ errors. The best fit Kepler frequencies (K)
for the resonance orbit ($\rm{r\sb{31}}$), the marginally stable orbit (ms) and at the event horizon (h) are compared
with the two high frequency measurements [4, 5]. Apparently Kepler frequencies for radii below the marginally stable orbit exist, 
at least over the duration of a flare}
\label{eps1}
\end{figure}

A closer look reveals that the average frequencies of the first three
groups are consistent with a frequency ratio of 1:2:3 (c.f. Fig. 1). 
Similar to what Abramowicz \& Klu\'zniak [6] have proposed to explain the 3:2 frequency ratio observed in microquasars 
I suggest that the 3:1 frequency ratio is due to a resonance between $\rm{\Omega\sb V}$ and $\rm{\Omega\sb R}$  
at some orbital radius r$\sb{31}$. 
The frequency in between is either the beat frequency of $\rm{\Omega\sb V}$ and $\rm{\Omega\sb R}$ or the 
first harmonic of $\rm{\Omega\sb R}$. This assumption results in a relation between r$\sb{31}$ and $a$. 
The additional requirement of the existence of a 3:2 resonance of $\rm{\Omega\sb V}$ and $\rm{\Omega\sb R}$ at a radius  r$\sb{32}$ and the same 
$a$ such that r$\sb{31}$ and r$\sb{32}$ are commensurable orbits, i.e. 
$\rm{\Omega\sb V}(\rm r\sb{31}) = 3\times \rm{\Omega\sb R}(\rm r\sb{32})$, produces a single solution for r$\sb{31}$ = 1.546, 
r$\sb{32}$ = 3.919 and $a = 0.99616$ [5]. r is measured in units of the gravitational radius. 
Interestingly, the same values of r$\sb{31}$ and $a$ can be derived in a totally different way. The inspection of the Boyer-Lindquist 
functions show that the orbital velocity $\rm{v\sp{(\Phi)}}$ described in the ZAMO-frame is no longer a monotonic function of r 
for $a > 0.9953$. In a small range of r  $\rm{\partial{v\sp{(\Phi)}}}/\rm{\partial r} > 0$.  
This is a new effect of General Relativity which has been overlooked so far. 
For  $2\pi{\rm{{{\partial{v\sp{(\Phi)}}}}}\over{\partial \rm r}} = \rm{\Omega\sb R}$ and 
$\rm{\Omega\sb V} = 3\times \rm{\Omega\sb R}$ the same values for 
r$\sb{31}$ and $a$ are obtained as above. With r$\sb{31}$ and $a$ fixed $\rm{M\sb{BH}}$ 
is given by just the observed frequencies, so that 
$\rm{M\sb{BH}/M\sb{\odot}} = 4603/\rm{\nu\sb{up}}$, with $\rm{\nu\sb{up}}$ the highest frequency of the triplet in Hz. 
For Sgr\,A* $\rm{M\sb{BH}} = (3.28\pm 0.13)\times 10\sp 6$ $\rm{M\sb{\odot}}$ and $a = 0.99616$ [5].  
This value of $\rm{M\sb{BH}}$ may be compared with the dynamically determined masses of 
$\rm{M\sb{BH}} = (3.59 \pm 0.59)\times$10$\sp 6$ $ \rm{M\sb{\odot}}$ [7], 
$ (4.07 \pm 0.62)\times$10$\sp 6$ $ \rm{M\sb{\odot}}$ [8] and 
$ (3.6 \pm 0.4)  \times$10$\sp 6$ $ \rm{M\sb{\odot}}$ [9]. 
The latter two measurements are for a distance of 8 kpc to Sgr\,A*.

\end{document}